\documentclass[pre,aps,preprint,showpacs]{revtex4-1}
\usepackage{revsymb4-1,latexsym,amssymb,amsmath,comment}
\usepackage{graphicx,psfrag,subfigure,stmaryrd,xcolor}
\usepackage[ansinew]{inputenc}

\newcommand{\be}{\begin{equation}}
\newcommand{\ee}{\end{equation}}
\newcommand{\bel}[1]{\begin{equation}\label{#1}}
\newcommand{\bea}{\begin{eqnarray}}
\newcommand{\eea}{\end{eqnarray}}
\newcommand{\ba}{\begin{array}}
\newcommand{\ea}{\end{array}}
\newcommand{\bef}{\begin{figure}}
\newcommand{\ef}{\end{figure}}

\begin{document}

\author{Thomas Bose and Steffen Trimper}
\affiliation{Institute of Physics,
Martin-Luther-University, D-06099 Halle, Germany}
\email{thomas.bose@physik.uni-halle.de}
\email{steffen.trimper@physik.uni-halle.de}
\title{Temperature gradient assisted magnetodynamics in a ferromagnetic nanowire}
\date{\today }

\begin{abstract}
The dynamics of the low energy excitations in a ferromagnet is studied in case a temperature gradient is 
coupled to the local magnetization. Due to the different time scales of changing temperature and magnetization 
it is argued that only the coupling between the spatially varying part of the temperature field and the 
magnetization is relevant. Using variational principles the evolution equation for the magnetic system is 
found which is strongly influenced by the local temperature profile. The system offers damped spin wave 
excitations where the strength of damping is determined by the magneto-thermal coupling. Applying the model to
nanowires it is demonstrated that the energy spectrum is significantly affected by the boundary conditions as 
well as the initial temperature distribution. In particular, the coupling between temperature and magnetization 
is expected to be several orders stronger for the open as for the isolated wire.


\end{abstract}

\maketitle

\section{Introduction}

Spin-based electronics is a topical and challenging subject in solid-state physics including the manipulation 
of spin degrees of freedom, in particular the spin dynamics and the spin-polarized transport
\cite{Zutic:RevModPhys76:323:2004}. In this regard it is desirable to find proper means for the creation and
control of spin currents. As it was firstly observed in \cite{Uchida:SSE:Nature:778:2008} a spin current can
be thermally generated by placing a magnetic metal in a temperature gradient. This phenomenon is called spin
Seebeck effect (SSE) and can be detected by attaching a Pt wire on top of the magnetic metal and measuring
a transverse electrical voltage in the Pt wire via the inverse Spin Hall effect, the conversion of a spin current
into an electric current \cite{Saitoh:ISHE:APL:182509:2006,Valenzuela:Nature442:176:2006}. Now a new subfield
has emerged called 'spin caloritronics' \cite{Bauer:SpinCal:SSC:459:2010} including the SSE as a local thermal
transport effect \cite{Sinova:SSE:NatMat9:880:2010}. In particular, the SSE combines spin degrees of freedom with
caloric properties. Further, the SSE was also observed in ferromagnetic semiconductors
\cite{Jaworski:SSEsc:NatMat9:898:2010}
and insulators \cite{Uchida:SSEinsul:NatMat9:894:2010}. Despite the absence of conduction electrons, a magnetic
insulator can convert heat flow into a spin voltage. The occurrence of a temperature-gradient induced spin current
along the magnetization direction in a ferromagnet is called longitudinal SSE
\cite{Uchida:LongitudSSEinsul:APL97:172505:2010}. Here, the magnon-induced spin current is injected parallel
to the temperature gradient from a ferromagnet into an attached paramagnetic metal.
Theoretically the interplay between temperature gradient and spin current within the SSE is investigated
intensively.  So, the SSE is attributed to a temperature difference between the magnon temperature in the
ferromagnet and the electron temperature in the attached Pt contact \cite{Xiao:MagnoDrivenSSE:PhysRevB81:214418:2010}.
Whereas in the conventional charge assisted Seebeck effect the Seebeck coefficient can be expressed by the
electronic conductivity, a microscopic model for spin transport driven by the temperature gradient was proposed
in \cite{Takazoe:SpinTempGrad:PhysRevB82:094451:2010}. The authors find thermally driven contributions to the spin
current in ferromagnetic metals. In case of ferromagnetic insulators a linear-response theory was
formulated \cite{Adachi:LinRespSSE:PhysRevB83:094410:2011} where the collective spin-wave excitations
are taken into account. A detailed numerical study of the SSE based on a modified Landau-Lifshitz-Gilbert equation
is presented in \cite{Ohe:NumSSE:PhysRevB83:115118:2011}. In addition, the scattering of a spin current
generated by a temperature gradient is analyzed in \cite{ChenglongBerakdar:SSE:PhysRevB83:180401:2011} where
the setup initiates a spin current torque acting on the magnetic scatterer. As pointed out in
\cite{Hinzke:PhysRevLett107:027205:2011} the domain wall motion can be also originated in magnonic spin currents.
Likewise it is stressed in \cite{Slonczewski:SpinTransferMagnons:PhysRevB82:0544032010} a spin-transfer torque in a
multilayer structure can be initiated by thermal transport from magnons. Recently
\cite{Adachi:EnhancedSSEPhonoDrag:APL97:252506:2010} reported that the SSE is considerably enhanced by nonequilibrium phonons
that drag the low-lying spin excitations. Due to a phonon-magnon drag mechanism one finds a phonon driven
spin distribution \cite{Jaworski:PhysRevLett106:186601:2011} which influences the SSE. The amplification of
magnons by thermal spin-transfer torque is observed in \cite{Pedron:AmplSpinWaves:PhysRevLett107:197203:2011}
where the system is subjected to a transverse temperature gradient. Another aspect of such collective
magnetic excitations in a magnetic insulator is the transmission of electrical signals
\cite{Kajiwara:nature464:262:2010}. Measurements of the Seebeck coefficient and the electrical resistivity of
polycrystalline magnetic films are presented in \cite{PhysRevB.83.100401}. The results provide important
information for further studies of the SSE in nanomagnets. Whereas most studies of spin caloric effects are
obtained for thin films, patterned ferromagnetic films demonstrate the profound effect of substrate
on the spin-assisted thermal transport \cite{PhysRevLett.107.216604}.

Motivated by recent experimental and theoretical findings related to SSE we study the magnetization dynamics
under the influence of a temperature gradient. Because the relevant time scale of the magnetization seems to be
totally different from the time scale of the temperature field our analysis is focused on the case that the
temperature gradient is preferably coupled to spatial variation of the magnetization. Due to spin-wave
excitations those spatial variations of the local magnetization have an impact on the spin dynamics.
To that aim we propose an action functional including both fields, the local magnetization field
$\mathbf{m}(\mathbf{x},t)$ as well as the local temperature field $T(\mathbf{x},t)$. The functional
includes a symmetry allowed coupling between the fields. Using variational principles
we derive the relevant Bloch-type equations of motion. The evolution equations are discussed in detail for one-dimensional
structures under distinct boundary conditions. In detail we study a ferromagnetic nanowire.

\section{Derivation of the equation of motion}

Let us consider a ferromagnet below its Curie temperature under the influence of a temperature field varying
in space and time. Then the action for the total system is composed of three parts
\begin{align}
\begin{aligned}
S=S_m +S_T +S_I \,.
\end{aligned}
\label{act1}
\end{align}
The magnetic part $S_m$ giving rise to the local precession of the magnetic moments $\mathbf{m}(\mathbf{x},t)$ reads \cite{BoseTrimper:PLA:2452:2011}
\begin{align}
\begin{aligned}
S_m=\int dt\int d^dx\,\left(\frac{J_{\alpha \beta}}{2}\,\frac{\partial m_\gamma}{\partial x_\alpha}\frac{\partial m_\gamma}{\partial x_\beta} + A_\alpha (\mathbf{m})\frac{\partial m_\alpha}{\partial t}\right) \,.
\end{aligned}
\label{actmag}
\end{align}
Here, the $J_{\alpha \beta}$ characterize the exchange interaction between the magnetic moments while the vector field $\mathbf{A}$ is a functional of the magnetic moments $\mathbf{m}$ as indicated in Eq.~\eqref{actmag}. It will be specified below. 
Because our system is confined in a finite volume it is appropriate to separate the scalar temperature field $T(\mathbf{x},t)=\Phi(\mathbf{x})\Theta(t)$. Following \cite{He:VariationHeatEq:PLA373:2614:2009} the temperature part is written as
\begin{align}
\begin{aligned}
S_T=&\int dt\int d^dx\,\left(\frac{\partial^2}{\partial x_\alpha^2} + \frac{1}{\kappa}\frac{\partial}{\partial t} + k^2\right)\,T(\mathbf{x},t)\\ =&\int dt\int d^dx\,\left(\Theta\,\frac{\partial^2\Phi}{\partial x_\alpha^2} + \frac{\Phi}{\kappa}\frac{\partial \Theta}{\partial t} + k^2 \Phi \Theta\right)\,.
\end{aligned}
\label{acttemp}
\end{align}
The constant $\kappa$ is the heat conductivity or temperature diffusivity while $k$ is the separation parameter
with the dimension of a wave vector. Its physical meaning will be discussed below.
The separation of the temperature field is likewise suggested by the well separated time scales of the
magnetization and the heat conduction. For that reason we argue that the interaction part $S_I$ includes only the coupling of the magnetization to the 
spatial varying part of the temperature field:
\begin{align}
\begin{aligned}
S_I=&\int dt\int d^dx\,\left(\Gamma_{\alpha \beta \gamma}\frac{\partial m_\alpha}{\partial x_\beta}\frac{\partial \Phi}{\partial x_\gamma} \right) \,.
\end{aligned}
\label{actint}
\end{align}
With other words the coupling reflects the assumption that the spatial variation of the magnetization is
directly coupled to the gradient of the $\mathbf{x}$-depending
part $\Phi (\mathbf{x})$. The tensor $\Gamma_{\alpha \beta \gamma}$ reflects the coupling strength.
As noticed above the separation parameter $k$ in Eq.~\eqref{acttemp} is proportional to an inverse length
scale $k \simeq l_{\text{\em \tiny T}}^{-1}$, describing the range where the temperature is significantly changed in space.
Having a nanowire or thin films in mind we assume $\mu m \leq l_{\text{\em \tiny T}} \leq mm$ defined by the length extensions of the sample. 
The characteristic time
$\tau_{\text{\em \tiny T}}$ for pure heat conductance in a confined medium is
$\tau_{\text{\em \tiny T}} \simeq (\kappa k^2)^{-1} \simeq \kappa^{-1} l^2_{\text{\em \tiny T}}$.
Taking the thermal diffusivity $\kappa =2.51\, \rm{mm}^2s^{-1}$ for Y$_3$Fe$_5$O$_{12}$ (YIG crystal)
\cite{Shen:YIGdata:MSEB:77:2009} one estimates the time scale $\tau_{\text{\em \tiny T}}$ in the range $\mu s \leq \tau_{\text{\em \tiny T}} \leq s$.
In contrast magnetodynamics occurs a considerably shorter time scale within the nanosecond range:
$\tau_m \leq ns$, i.e. $\tau_{\text{\em \tiny T}}$ is supposed to differ from $\tau_m$ about a few orders of magnitude,
$\tau_m \ll \tau_{\text{\em \tiny T}}$. From Eqs.~\eqref{act1}-\eqref{actint} one finds the equations of motion by
variational principles:
\begin{subequations}
\begin{align}
\frac{\partial^2\Phi}{\partial x_\alpha^2} + k^2\Phi &= 0 \,,\label{deltaTheta}\\
\frac{1}{\kappa}\frac{\partial \Theta}{\partial t} + k^2\Theta &= \Gamma_{\alpha \beta \gamma }\,
\frac{\partial ^2 m_\alpha}{\partial x_\beta \partial x_\gamma }
\,,\label{deltaPhi}\\
\left( \frac{\partial A_\beta }{\partial m_\alpha } -  \frac{\partial A_\alpha }{\partial m_\beta }\right) \,
\frac{\partial m_\beta }{\partial t} &= J_{\beta \gamma }\,\frac{\partial ^2 m_\alpha }
{\partial x_\beta \partial_\gamma } + \Gamma_{\alpha \beta \gamma }\, \frac{\partial ^2\Phi}{\partial x_\beta
\partial x_\gamma} \,.\label{deltam}
\end{align}
\label{delta}
\end{subequations}
This set of equations describes the influence of a temperature profile on the magnetization, Eq.~\eqref{deltam},
and the feedback of the magnetization on the temperature, Eq.~\eqref{deltaPhi}. In a first approximation we
neglect the latter one. Consequently, Eqs.~\eqref{deltaTheta} and \eqref{deltaPhi} represent the conventional
heat equation after the separation of the variables. In the following we consider an isotropic magnet
characterized by $J_{\alpha \beta}= J\,\delta_{\alpha \beta}$, see Eq.~\eqref{deltam}. Multiplying
Eq.~\eqref{deltam} with $\mathbf{m}$ vectorial it results
\begin{align}
\begin{aligned}
\dot{m}_\alpha\,  \left( \mathbf{m} \cdot\mathbf{curl_m}\mathbf{A}\right) = J\,
\left(\mathbf{m}\times \nabla^2\mathbf{m}\right)_\alpha  + \Lambda_{\alpha \beta \gamma \delta }\,m_\beta \,
\frac{\partial^2 \Phi}{\partial x_\gamma  \partial x_\delta }\,.
\end{aligned}
\label{deltam2}
\end{align}
Here there appears a tensor $\Lambda_{\alpha \beta \gamma \delta }$, which includes the coupling between
magnetization and temperature profile according to Eq.~\eqref{actint}. Since $\boldsymbol\Lambda$ is symmetric
with respect to the indices $\gamma$ and $\delta$ one gets for isotropic systems
$\Lambda_{\alpha \beta \gamma \delta } = C_1 \delta_{\alpha \beta }\delta_{\gamma \delta } +
C_2 ( \delta_{\alpha \gamma } \delta_{\beta \delta } + \delta_{\alpha \delta} \delta_{\beta \gamma})$.
If $\boldsymbol\Lambda =0$ Eq.~\eqref{deltam2} should conserve the spin length
$\mathbf{m}^2 = 1$ because of the absence of dissipative forces.
From here we conclude that a pure reversible equation of motion can be obtained by
setting $\mathbf{curl_m}\mathbf{A}=-\gamma^{-1}\,\mathbf{m}$ where $\gamma$ is the gyromagnetic ratio.
As the result one gets the Landau-Lifshitz equation without relaxation. Using the same ansatz also for
nonzero $\boldsymbol\Lambda$ and setting in leading order $C_1 = - C$ and $C_2 = 0$
one finds from Eq.~\eqref{deltam2} the following equation
\begin{align}
\begin{aligned}
\frac{\partial \mathbf{m}(\mathbf{x},t)}{\partial t} = -\gamma \,J\,\mathbf{m}(\mathbf{x},t)\times
\nabla^2\mathbf{m}(\mathbf{x},t) + \gamma \,C\, \left[ \nabla^2 \Phi(\mathbf{x})\right]\,\mathbf{m}(\mathbf{x},t) \,.
\end{aligned}
\label{eommag}
\end{align}
The sign of the coupling parameter $C$ between temperature and magnetization will be discussed in detail below.
The final evolution equation Eq.~\eqref{eommag} does not conserve the spin length anymore and has the form of
a Bloch-Bloembergen-like equation \cite{Bloch:PhysRev70:460:1946,Bloembergen:PhysRev78:572:1950}. Therefore it
should be applicable for ferromagnetic systems on a mesoscopic scale.
To affirm the reasonability of the result in Eq.~\eqref{eommag} we briefly outline an alternative way of
its derivation. Now the starting point is the
energy functional of the total system 
\begin{align}
\begin{aligned}
\mathcal{F}=\mathcal{F}_m + \mathcal{F}_T + \mathcal{F}_I \,,
\end{aligned}
\label{enfunc}
\end{align}
where it is likewise decomposed into a magnetic part $\mathcal{F}_m$, a temperature part $\mathcal{F}_T$ and an interaction
part $\mathcal{F}_I$.
Within the Ginzburg-Landau theory the functional is given by
\begin{align}
\begin{aligned}
\mathcal{F}=\int d^dx\, \left( \frac{J_{\alpha \beta}}{2}\,\frac{\partial m_\gamma}{\partial x_\alpha}\frac{\partial m_\gamma}{\partial x_\beta} + \frac{\chi_{\alpha \beta}}{2}\,\frac{\partial T}{\partial x_\alpha}\frac{\partial T}{\partial x_\beta} + \Gamma_{\alpha \beta \gamma}\,\frac{\partial m_\alpha}{\partial x_\beta}\frac{\partial \Phi}{\partial x_\gamma}\right) \,,
\end{aligned}
\label{enfuncexpl}
\end{align}
where the coupling tensors $J_{\alpha \beta}$ and $\Gamma_{\alpha \beta \gamma}$ are the same as before in Eqs.~\eqref{actmag} and \eqref{actint}.
Moreover, we have chosen a quadratic coupling of the gradient of the temperature characterized by
$\chi_{\alpha \beta}$. The function $\Phi =\Phi (\mathbf{x})$ is
again the space-dependent part of the scalar temperature field $T(\mathbf{x},t)=\Phi (\mathbf{x})\Theta (t)$.
In order to derive the desired equations of motion we follow the common approach \cite{ChaikinLubensky2000principles}
\begin{subequations}
\begin{align}
\frac{\partial T}{\partial t} &= -\rho_{\text{\em \tiny T}}\,\frac{\delta \mathcal{F}}{\delta T} \,,\label{tempGL}\\
\frac{\partial m_\alpha }{\partial t} &= -\rho_m\, \int d^dy\,\left\{ m_\alpha  (x),m_\beta  (y)\right\}\,\frac{\delta \mathcal{F}}{\delta {m_\beta  (y)}} =
\rho_m\,\epsilon_{\alpha \beta \gamma }m_\beta \frac{\delta \mathcal{F}}{\delta m_\gamma } \label{magGL}\,.
\end{align}
\label{eomGL}
\end{subequations}
Here, the Poisson brackets $\{m_\alpha (x),m_\beta (y)\}=\delta \,(x-y)\,\epsilon_{\alpha \beta \gamma }m_\gamma (x)$ 
for classical magnetization densities are introduced, see \cite{book:Mazenko:Nonequilibrium:2006,ChaikinLubensky2000principles}.
Now we use the separation ansatz for the temperature in Eq.~\eqref{tempGL} and restrict the calculation to isotropic situations;
$J_{\alpha \beta}=J\,\delta_{\alpha \beta}$, $\Lambda_{\alpha \beta \gamma \delta }= -C\,\delta_{\alpha \beta}\delta_{\gamma \delta }$
and additionally $\chi_{\alpha \beta}=\chi \,\delta_{\alpha \beta}$.
Identifying further $\chi \rho_{\text{\em \tiny T}}=\kappa$ and $\rho_m=-\gamma$ the resulting evolution equations coincide exactly with 
Eqs.~\eqref{deltaTheta}, \eqref{deltaPhi} and \eqref{eommag} for the temperature and the magnetization, respectively, 
as derived from the expressions for the action in Eqs.~\eqref{act1}-\eqref{actint}.
The advantage of the second method is that the firstly unknown quantity $\mathbf{A}(\mathbf{m})$, introduced in Eq.~\eqref{actmag}, is not necessary. On the other hand, the
coincidence of the obtained results for the equations of motion based on different approaches justifies the assumptions made for the vector field $\mathbf{A}$.

\section{Application to ferromagnetic nanowires}

In this section Eq.~\eqref{eommag} is analyzed for a ferromagnetic nanowire as schematically depicted in
Fig.~\ref{picwire}.
\bef
\includegraphics[width=8cm]{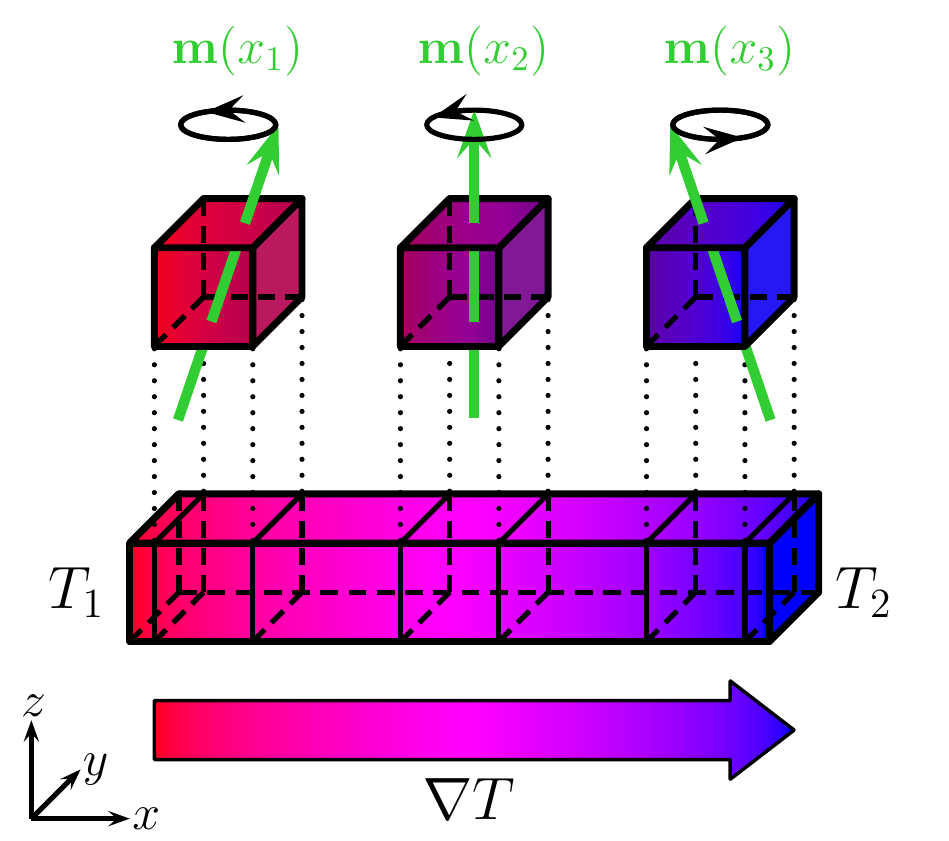}
\caption{(Color online) Nanowire with locally varying temperature.}
\label{picwire}
\ef
Because the model is focused on the ferromagnetic phase and the related collective low energetic excitations, the magnetization-field is decomposed 
into
$\mathbf{m}(x,t)= \mathbf{m}_0 + \boldsymbol\psi (x,t) \equiv \psi_1(x,t) \,\mathbf{e}_1 +
\psi_2(x,t)\,\mathbf{e}_2 + m_0\,\mathbf{e}_3$. Here $m_0$ is the constant magnetization whereas the spin-wave amplitudes $\psi_1, \psi_2 $ obey the coupled equations
\begin{align}
\begin{aligned}
\frac{\partial \psi_1}{\partial t} &= \gamma\,J\,m_0\,\frac{\partial ^2\psi_2}{\partial x^2} + \gamma\,C\,\frac{\partial ^2\Phi}{\partial x^2}\,\psi_1 \,,\\
\frac{\partial \psi_2}{\partial t} &= -\gamma\,J\,m_0\,\frac{\partial ^2\psi_1}{\partial x^2} + \gamma\,C\,\frac{\partial ^2\Phi}{\partial x^2}\,\psi_2 \,.
\end{aligned}
\label{magsys}
\end{align}
The magnetization dynamics is accordingly influenced only by the spatially varying temperature field $ \Phi(x)$, the complete dynamics of
which are given by Eqs.~\eqref{deltaTheta} and Eq.~\eqref{deltaPhi}. In one dimension it results
\begin{align}
\frac{\partial ^2\Phi}{\partial x^2}=-k^2\,\Phi\,,\quad
\frac{\partial \Theta}{\partial t}=-\kappa \,k^2\,\Theta \label{eomPhi} \,.
\end{align}
The solution for $\Phi(x)$ in Eq.~\eqref{eomPhi} can be written as a superposition $\Phi(x) =\sum_n \Phi_n(x) $. Hence
$\partial ^2\Phi/\partial x^2=-\sum_n\,k_n^2\Phi_n$ can be substituted in Eq.~\eqref{magsys}.
In terms of the Fourier transformed $\tilde {\Phi}_n(q,t)=\mathcal{FT}\{\Phi (x,t)\}$ and $\varphi_i (q,t)=\mathcal{FT}\{\psi_i (x,t)\}$ Eq.~\eqref{magsys}
can then be transformed into
\begin{align}
\begin{aligned}
\frac{\partial}{\partial t}\varphi_1(q,t) &= -a^2\,q^2\,\varphi_2(q,t) - r\,a^2\,\sum_n\left(k_n^2\,\int dq'\,\varphi_1(q-q',t)\,\tilde{\Phi}_n(q')\right) \,,\\
\frac{\partial}{\partial t}\varphi_2(q,t) &= a^2\,q^2\,\varphi_1(q,t) - r\,a^2\,\sum_n\left(k_n^2\,\int dq'\,\varphi_2(q-q',t)\,\tilde{\Phi}_n(q')\right) \,.
\end{aligned}
\label{magsysFT}
\end{align}
Here the time $t$ is measured in terms of $\bar{t}= \gamma \,\tilde{J}\,m_0\,t $, where for convenience
$\bar{t}$ is already called $t$ in Eq.~\eqref{magsysFT}. The other dimensionless quantities are defined by
\begin{gather}
\begin{gathered}
J=\tilde{J}\,a^2\,,\,C=K\,m_0\,a^2\,,\,r=K/\tilde{J}\,,
\end{gathered}
\label{scaling}
\end{gather}
where $a$ is the lattice constant. The separation parameters $k_n$ and the $\tilde{\Phi}_n(q)$ are determined
as the solution of Eq.~\eqref{eomPhi} under consideration of the relevant boundary conditions.

\subsection{Zero heat flow at the boundaries}

Let us firstly study the case when there is no heat current flowing through the edges of the wires. The initial and boundary conditions for the heat equation 
Eq.~\eqref{eomPhi} read
\begin{align}
\begin{aligned}
T(x,t=0) &= \Theta_0\sum_n\Phi_n(x) \,,\quad \left.\frac{\partial T(x,t)}{\partial x}\right|_{x=0,L} &= 0\,,
\end{aligned}
\label{icbc1}
\end{align}
where $L$ is the length of the nanowire.
The solution of Eq.~\eqref{eomPhi} fulfilling the boundary condition Eq.~\eqref{icbc1} is
\begin{align}
\begin{aligned}
\Phi_n(x)=u_n\,\cos(k_n\,x) \,,\qquad k_n=\frac{n\,\pi}{L} \,,\quad n=0,1,2,...\,.
\end{aligned}
\label{solPhi}
\end{align}
Performing the Fourier transformation we obtain
\begin{align}
\begin{aligned}
\tilde{\Phi}_n(q)=u_n\,\pi\,\left[\delta\,(q+k_n)+\delta\,(q-k_n)\right] \,.
\end{aligned}
\label{FTPhi}
\end{align}
Inserting this relation into Eq.~\eqref{magsysFT} leads to terms of the form $\varphi_i(q\pm k_n)$ which can be
expanded with respect to the small parameters $k_n$:
\begin{align}
\begin{aligned}
\varphi_i(q\pm k_n) \simeq \varphi_i(q) \pm \frac{\partial \varphi_i}{\partial q}\,k_n + \mathcal{O}\left(k_n^2\right)\,.
\end{aligned}
\label{expansionphi}
\end{align}
Is $\lambda$ the wave length of the low energetic excitation in the ferromagnet with the corresponding wave vector
$q \sim 1/\lambda$ and setting $k_n\sim 1/L$ the expansion made in Eq.~\eqref{expansionphi} is valid for
$k_n\ll q\ll 1/a$ or $L\gg \lambda \gg a$, respectively.
Introducing $\eta=\varphi_1 + \mathrm{i}\,\varphi_2$ and combining Eqs.~\eqref{FTPhi}-\eqref{expansionphi} with Eq.~\eqref{magsysFT} one obtains the following evolution
equation for $\eta$:
$$
\frac{\partial}{\partial \bar{t}}\eta(q,\bar{t}) =
\left(\mathrm{i}\,a^2q^2 - \bar{\alpha}\right)\,\eta(q,\bar{t})\,,\quad \bar{\alpha} =2\pi r a^2\,\sum_n k_n^2\,u_n \,.
$$
The solution for the $\boldsymbol\varphi=(\varphi_1,\varphi_2)$ with the initial conditions $\boldsymbol\varphi(t=0)=\boldsymbol\varphi_0=(\varphi_{01},\varphi_{02})$ 
is written in terms of the real time $t$ as
\begin{align}
\begin{aligned}
\boldsymbol\varphi(q,t) = \begin{pmatrix}
            \cos\left[\omega(q)\,t\right] & -\sin\left[\omega(q)\,t\right]\\
            \sin\left[\omega(q)\,t\right] &  \cos\left[\omega(q)\,t\right] \end{pmatrix}\,\boldsymbol\varphi_0\,\exp[-\alpha\,t] \,,\qquad \omega(q)= \gamma\,J\,m_0 a^2q^2 \,.
\end{aligned}
\label{sol1}
\end{align}
Here the spin wave energy $\omega(q)$ and the damping constant $\alpha$ are defined by
\begin{align}
\begin{aligned}
\omega(q)= \gamma\,J\,m_0 q^2 \,,\quad \alpha = 2\pi r J  \gamma m_0 \sum_n k_n^2\,u_n \,.
\end{aligned}
\label{disdamp}
\end{align}
The expression for $\omega(q)$ reflects the common quadratic spin wave dispersion relation.
The temperature gradient leads to damped spin waves whereas the damping parameter $\alpha$ is defined in
Eq.~\eqref{disdamp}. The strength of the attenuation $\alpha $ can be estimated from
Eq.~\eqref{disdamp} and Eq.~\eqref{scaling} using $\alpha \propto C \sum_nk_n^2\,u_n$. In order that the
damping is physically realized we have to discuss the sign of the parameter $C \propto r$ introduced in
\eqref{eommag}. It characterizes the coupling between the local magnetization and the local temperature.
Moreover, the sign of $\sum_nk_n^2\,u_n$ is determined by the sign of the coefficients $u_n$
which is given in terms of the initial temperature distribution, compare
Eqs.~\eqref{icbc1}-\eqref{solPhi}. A positive damping parameter $\alpha > 0 $ is realized by assigning
\begin{align}
\sum_nk_n^2\,u_n \lessgtr 0 \quad \Longrightarrow \quad r\lessgtr 0 \,.
\label{casesC1}
\end{align}
The $t$- and $q$-dependent ratio $\varphi_1/\varphi_0$ is plotted in Fig.~\ref{figsw3dcos}.
\bef
\centering
\subfigure[]{%
				\label{figsw3dcos}
				\includegraphics[width=8cm]{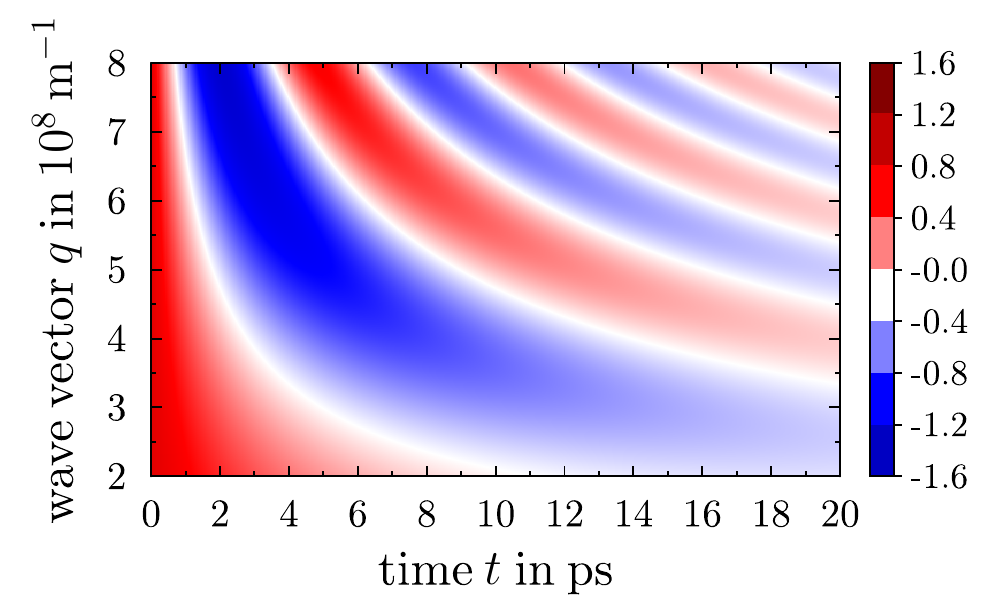}}\\
\subfigure[]{%
				\label{figsw3dsin}
				\includegraphics[width=8cm]{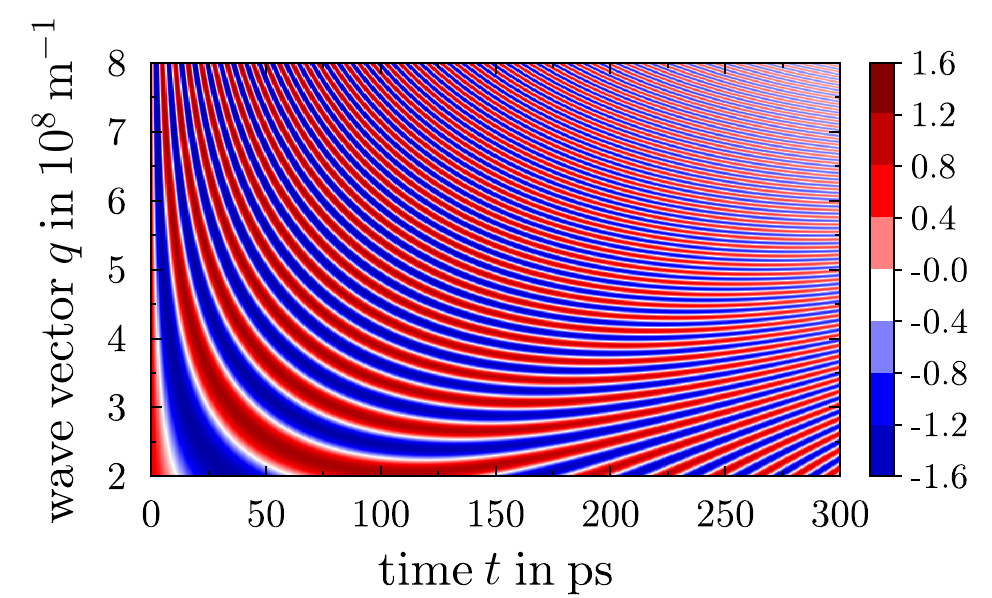}}
\caption{(Color online) Spin waves $\varphi_1(q,t)/\varphi_0$ (color scale) for $a=1\mathrm{\AA}$, $\mathrm{L}=1\mu \mathrm{m}$ and $r=100$ for the cases: (a)~vanishing heat current at the edges of the nanowire according to the initial heat profile I in the inset in Fig.~\ref{figsolcos};
(b)~fixed temperatures at the edges referring to the initial heat profile III in the inset in Fig.~\ref{figsolsin}.}
\label{figsw3d}
\ef
As expected the frequency is increased when $q$ is enlarged. Moreover the damping should be observed within a picoseconds time-scale. To illustrate that we present the dynamic solutions for different heat profiles in Fig.~\ref{figsolcos}.
\bef
\centering
\subfigure[]{%
				\label{figsolcos}
				\includegraphics[width=8cm]{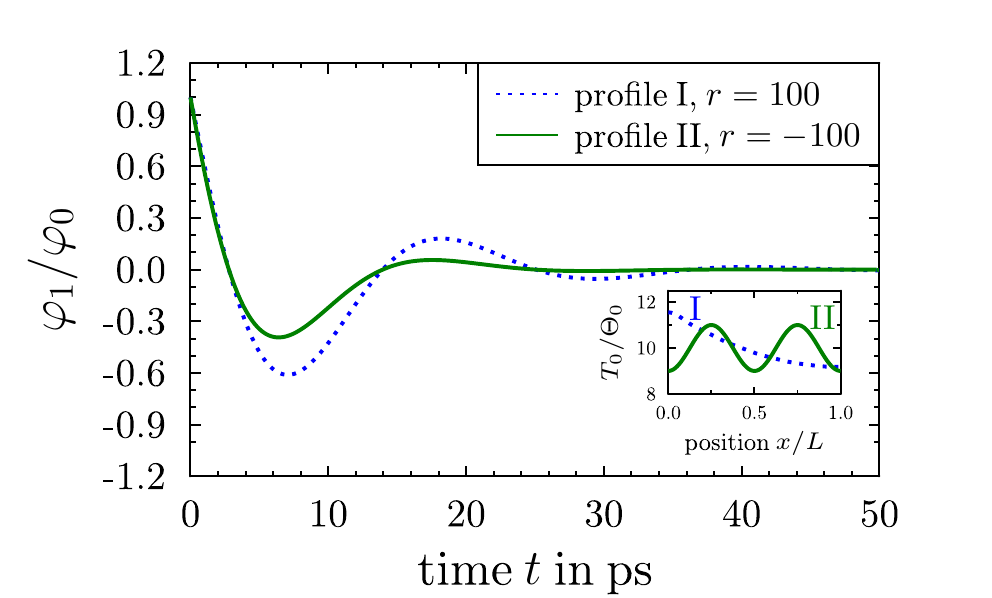}}\\
\subfigure[]{%
				\label{figsolsin}
				\includegraphics[width=8cm]{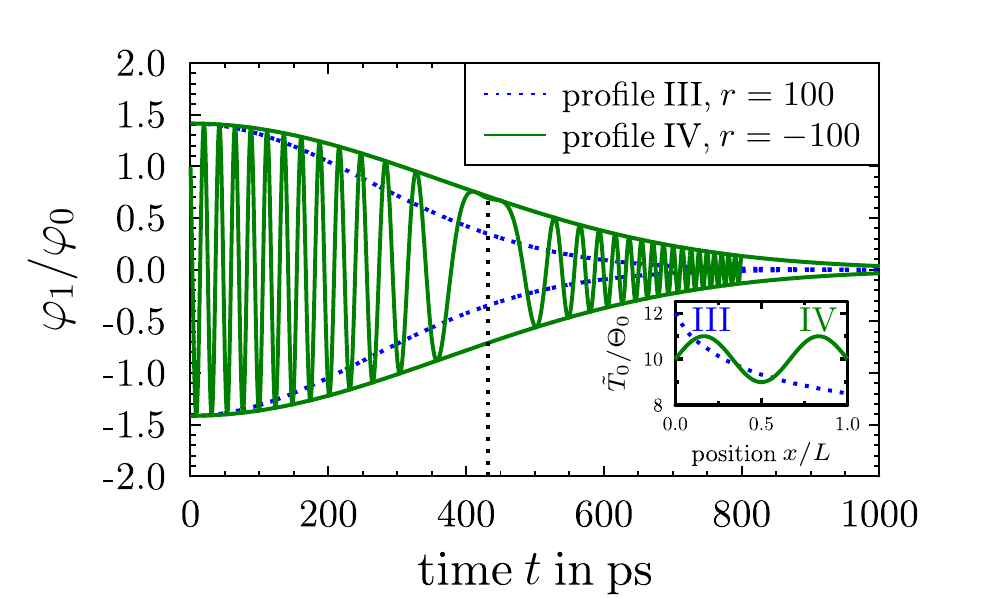}}
\caption{(Color online) Spin wave solutions for $a=1\mathrm{\AA}$, $\mathrm{L}=1\mu \mathrm{m}$ and $q=4\times 10^8\textrm{m}^{-1}$ for the cases: (a)~vanishing heat current at the edges of the nanowire with the initial heat profiles I ($u_n=1/n^2$, $u_0=10$) and II ($u_n=u_4=-1$, $u_0=10$) mentioned in the text;
(b)~fixed temperatures at the edges with the initial heat profiles III ($v_n=-1/n^2$, $T_1/\Theta_0=12$, $T_2/\Theta_0=8.5$) and IV ($v_n=v_3$, $T_1/\Theta_0=10=T_2/\Theta_0$) in the inset mentioned in the text. Here only the envelope of the solution referring to heat profile III is plotted.}
\label{figsol}
\ef
The initial temperature distribution is chosen by $T_0/\Theta_0=u_0+\sum_n u_n\,\cos[k_n\,x]$ and is depicted in the inset of Fig.~\ref{figsolcos}.
To reach a damping of the spin wave in the ps-range we set $|r|= K/\tilde{J} = 100$, where $r$ is a measure for the relation of the coupling of the temperature and magnetization profiles
(i.e. $\nabla\Phi\nabla\mathbf{m}$) and the magnetization-magnetization (i.e. $\nabla\mathbf{m}\nabla\mathbf{m}$) coupling, see Eq.~\eqref{scaling}. In that case one observes a strong coupling between temperature and magnetization.

\subsection{Fixed temperatures at the boundaries}

As the second realization we consider the case that the edges of the wire are held at fixed temperatures:
\begin{align}
\begin{aligned}
T(x,t=0) &=& \Theta_0\sum_n\Phi_n(x) \,,\quad \Phi(x = 0) = \Phi(x=L) = 0 \,.
\end{aligned}
\label{icbc2}
\end{align}
Therefore, Eq.~\eqref{icbc2} refers to the transformed temperature field $T(x,t)=\tilde{T}(x,t)-(1-x/L)\,T_1-x/L\,T_2$, where $T_1=\tilde{T}(x=0,t)$ and $T_2=\tilde{T}(x=L,t)$
are fixed and $\tilde{T}(x,t)$ is the original temperature profile in the wire.
The spatial temperature profile $\Phi(x)$ is obtained from Eq.~\eqref{eomPhi} combined with the boundary
conditions in Eq.~\eqref{icbc2}. The solution is the superposition of
\begin{align}
\begin{aligned}
\Phi_n(x)=v_n\,\sin\left(k_n\,x\right) \,,\qquad k_n=\frac{n\,\pi}{L} \,,\quad n=1,2,... \,.
\end{aligned}
\label{solPhi2}
\end{align}
The Fourier transform of that $\Phi_n(x)$ reads
\begin{align}
\begin{aligned}
\tilde{\Phi}_n(q)=\mathrm{i}\,v_n\,\pi\,\left[\delta\,(q+k_n)-\delta\,(q-k_n)\right] \,.
\end{aligned}
\label{FTPhi2}
\end{align}
Utilizing again the expansion in Eq.~\eqref{expansionphi} and the transformation $\eta=\varphi_1 + \mathrm{i}\,\varphi_2$ we find by inserting Eq.~\eqref{FTPhi2} in Eq.~\eqref{magsysFT}
\begin{align}
\begin{aligned}
\frac{\partial}{\partial\bar{t}}\eta(q,\bar{t}) + \mathrm{i}\,\bar{\beta}\,\frac{\partial}{\partial q}\eta(q,\bar{t}) = \mathrm{i}\,a^2q^2\,\eta(q,\bar{t}) \quad ,\qquad \bar{\beta} = 2\pi r a^2\,\sum_n k_n^3\,v_n \,.
\end{aligned}
\label{eta2}
\end{align}
This partial differential equation of first order can be solved by using the method of characteristics.
The solution for $\boldsymbol\varphi=(\varphi_1,\varphi_2)$ with the initial conditions $\boldsymbol\varphi(t=0)=\boldsymbol\varphi_0$ is given by
\begin{align}
\begin{aligned}
\boldsymbol\varphi(q,t) = \begin{pmatrix}
            \cos\left[\Omega(q,t)\,t\right] & -\sin\left[\Omega(q,t)\,t\right]\\
            \sin\left[\Omega(q,t)\,t\right] &  \cos\left[\Omega(q,t)\,t\right] \end{pmatrix}\,\boldsymbol\varphi_0\,\exp[J \gamma m_0 \beta(t) q\,t] \,.
\end{aligned}
\label{sol2}
\end{align}
The quadratic spin wave dispersion relation $\Omega(q,t)$ and the damping parameter $\beta(t)$ are defined by
\begin{align}
\begin{aligned}
\Omega(q,t)=J \gamma m_0 \left(\,q^2 - \frac{\beta^2(t)}{3}\,\right)\,,\qquad \beta(t) =  2\pi r J \gamma m_0\,t \,\sum_n k_n^3\,v_n\,.
\end{aligned}
\label{dampdis}
\end{align}
Obviously the quadratic spin wave dispersion relation is now modified by a time depending correction term
$J \gamma m_0 \beta^2(t)/3$, i.e. the spin wave dispersion is changed by the damping parameter $\beta(t)$ which
occurs as a factor in the exponential function in Eq.~\eqref{sol2}.
The decay of this function is guaranteed for negative values $\beta(t) <0$.
In the same manner as before, using Eq.~\eqref{dampdis} and Eq.~\eqref{scaling}, we estimate
$\beta \propto C\sum_nk_n^3\,v_n$. Thus, one has to distinguish
\begin{align}
\sum_nk_n^3\,v_n \lessgtr 0 \quad \Longrightarrow \quad r\gtrless 0 \,.
\label{casesC2}
\end{align}
Referring to the initial heat distribution ($\tilde{T}$ was introduced below in Eq.~\eqref{icbc2}) we assume
$\tilde{T}_0/\Theta_0=(1-x/L)T_1/\Theta_0+xT_2/L\Theta_0+\sum v_n\,\sin[k_n\,x]$, see the inset in Fig.~\ref{figsolsin}.
The solution $\varphi_1(q,t)/\varphi_0$ is shown in Fig.~\ref{figsw3dsin}. Contrary to the previous case
of an absent heat flow through the ends of the wire now we observe an enhancement of the lifetime of the spin waves,
cf. the time scales of Figs.~\ref{figsw3dcos} and \ref{figsw3dsin}. 
The same effect of increased spin wave lifetimes is noticeable in Fig.~\ref{figsol} where the solution for $q=4\times 10^8\textrm{m}^{-1}$ is 
illustrated for different heat profiles but identical values of the coupling relation $r$ for both types of boundary conditions. 
In Fig.~\ref{figsolsin} we only plot the envelope of the solution for one of both cases.
However, the curve shown is featured by a decrease of the frequency up to a time of $\sim 430\textrm{ps}$ 
which is determined by the maximum of the argument $\Omega (q,t)t$ of the periodic functions. Thereafter the frequency increases 
continuously with time. 
A quantitative comparison of the lifetimes for comparable temperature distributions in Fig.~\ref{figsol} offers that the characteristic time scale 
is $40-90$ times larger for the case when heat flows through the edges than for the case the nanowire is isolated.
The reason for that consists of the expansion made in Eq.~\eqref{expansionphi}. Because in Eq.~\eqref{eta2} all even terms cancel mutually in the combination $\varphi_i(q+k_n)-\varphi_i(q-k_n)$ and hence only the odd terms remain.
This leads to an alteration of the damping constant in Eqs.~\eqref{sol2}-\eqref{dampdis} compared to $\alpha$ in Eqs.~\eqref{sol1}-\eqref{disdamp}.
The physical picture behind might be as follows: As discussed the type of the temperature profile $\Phi(x)$ determines the 
strength of coupling. In our model the solution for $\Phi(x)$ is selected by the boundary conditions either in Eq.~\eqref{icbc1} or in Eq.~\eqref{icbc2}. 
The complete isolation of the nanowire imposes the restriction on the thermo-magnetic system that it is not able to exchange energy with the environment. This leads to 
very limited possibilities for the self-organization of the magneto-caloric behavior. Otherwise, if heat is allowed to flow through the edges of the wire the 
thermo-magnetic system is in contact with its surrounding. Thus the self-organization process can be optimized due to a broader variety of its realization.

\section{Conclusions}

In the present work the low energetic excitations in a ferromagnetic nanowire had been analyzed when the spatial variations of the
magnetization is coupled to the corresponding temperature gradient. The resulting spin-waves are damped due to the temperature gradient. 
This result contributes to the recent effort in understanding spin caloritronics phenomena. Whereas the spin Seebeck effect is focused on 
the creation of a spin current due to a temperature gradient the goal of our study is to elucidate the behavior of the permanent effect of that temperature gradient 
on the excitation of the magnetic system. Once the magnetic system is excited the shape of the temperature profile influences the damping, too. Further, 
the external conditions under which an experiment is realized should affect the dynamics as well. In this regard we found that the
system is strongly controlled by the boundary conditions which modify the damping parameter. While the complete isolated nanowire offers conventional spin wave 
excitations which are damped due to the coupling to a spatial changing temperature field the situation is more complex in case the ends of the nanowire are held at fixed 
temperatures. Referring to the latter realization it results that the damping parameter is time-dependent and modifies the spin wave dispersion relation. 
However, if one considers systems which are subjected to identical boundary conditions the actual initial temperature distribution also affects the spin dynamics.
In principle it should be possible to validate our theoretical findings experimentally by spin Seebeck effect measurements or ferromagnetic resonance techniques in 
low dimensional samples. Our studies can be accordingly extended to higher dimensions. Further, the model can be refined by taking into account higher order terms in the 
expansion made in Eq.~\eqref{expansionphi}. The inclusion of stochastic forces should also improve the model slightly.\\

One of us (T.B.) is grateful to the Research Network 'Nanostructured
Materials'\,, which is supported by the Saxony-Anhalt State, Germany.

\clearpage

\bibliography{Tgrad}
\bibliographystyle{apsrev4-1}

\end{document}